\documentstyle[12pt,fullpage]{article}

\def\rarrow{\:\raisebox{-1.3ex}{$\stackrel{\longrightarrow}{t\rightarrow\infty}$}\:}

\begin{document}
\begin{flushright}
TIFR/TH/99-46
\end{flushright}
\bigskip
\begin{center}
{\bf Exact Solutions of the Caldeira-Leggett Master Equation: 
A Factorization Theorem For Decoherence}\\[1cm] 
S. M. Roy$^{\dagger,\star}$ and Anu
Venugopalan$^{\dagger\dagger,\star\star}$\\[1cm] 
$^\dagger$Department of Theoretical Physics, Tata Institute of 
Fundamental Research \\
Mumbai - 400 005, INDIA  \\[1cm]
$^{\dagger\dagger}$Physical Resarch Laboratory, 
Navrangpura,  \\ Ahmedabad - 380 009, INDIA   
\end{center}

\begin{abstract}
Exact solutions of the Caldeira-Leggett  Master equation 
for the reduced
density matrix $\langle x'|\rho(t)| x \rangle$ for a free particle
and for a harmonic oscillator system coupled to a heat bath of
oscillators  are obtained for arbitrary initial conditions. The 
solutions prove that the Fourier transform of  $\rho(t)$ 
with respect to $\frac{(x + x')}{2}$
factorizes exactly into a part depending linearly on $\rho(0)$ and 
a part independent of it. The theorem yields the exact initial state dependence of
$\rho(t)$ and its eventual diagonalization in the energy basis.
\end{abstract}

\vfill
\vfill

\noindent $^\star$E-mail: {\em shasanka@theory.tifr.res.in}

\noindent $^{\star\star}$E-mail: {\em anu@prl.ernet.in}

\newpage

\noindent {\bf I. Introduction --} Feynman and Vernon\cite{fey} pioneered the
`influence functional technique' 
in which a quantum system of interest and its environment are represented 
by a single
Hamiltonian and the time development of the reduced density matrix $\rho$ of
the system is computed by tracing out the degrees of freedom of the 
environment. Using this technique on the total Hamiltonian
\begin{equation}
H=\frac{p^{2}}{2m} + V(x) + x \sum_{k}{}c_{k}R_{k} + 
\sum_{k} \Big ( \frac{p_{k}^{2}}{2M} + \frac{1}{2}M 
\omega_{k}^{2}R_{k}^{2} \Big ),
\end{equation}
in which the environment is modeled as a thermal bath of oscillators with 
co-ordinate co-ordinate coupling to the system coordinate $x$, 
Caldeira and Leggett\cite{cl} derived  the following high-temperature
master equation for the reduced density operator:
\begin{equation}
\frac{\partial{\rho}}{\partial{t}} = -\frac{i}{\hbar}\Big [H_{R},\rho \Big ] -
\frac{i\gamma}{\hbar} \Big ( \Big [\frac{1}{2}\{p,x\}, \rho \Big ] 
+ \Big[ x, \rho p \Big ]
- \Big [p, \rho x \Big ] \Big ) - \frac{D}{4 \hbar^{2}}
\Big [x,\Big [x,\rho \Big
] \Big],
\end{equation}
where $H_{R}$ is the renormalized system Hamiltonian, 
$\gamma$ is the relaxation rate and 
\begin{equation}
D=8m\gamma k_{B}T,
\end{equation}
where $k_{B}$ is the Boltzmann constant and $T$ is the temperature of the 
bath.
The same master equation has also been derived from completely 
different approaches and approximations\cite{dek,gsa}. In particular, for
\begin{equation}
H_{R}= \frac{p^{2}}{2m} + \frac{1}{2}m\omega^{2}x^{2},
\end{equation}
Agarwal's approximations yield Eq(2) with
\begin{equation}
D=8m\gamma \hbar \omega(\bar{n} + \frac{1}{2}),
\end{equation}
where 
\begin{equation}
\bar{n}=\Big [ \exp \left({\hbar \omega \over k_{B}T}\right) - 1 \Big ]^{-1},
\end{equation}
which has been used even at zero temperature by some authors\cite{sw}.
In that case, at all temperatures,
\begin{equation}
D \ge 4m\gamma \hbar \omega.
\end{equation}
For an open system, coupling with this environment may lead to near 
diagonalization of the reduced density matrix in some preferred basis.
Such decoherence has obvious conceptual interest in understanding 
the transition from quantum to classical behaviour\cite{zeh}. 
In addition, advances of technology have increased the propsects of tests
of decoherence by producing superpositions of macroscopically 
distinguishable states\cite{brune}. 
On the other hand, maintenance of quantum 
coherence is crucial to the success of 
quantum computation, cryptography and 
teleportation\cite{shor}.

The purpose of the present work is a rigorous study of quantum 
decoherence by means of exact solutions of the Caldeira-Leggett equation. 
Some interesting questions which have been stimulated by the work of Zurek,
Habib and Paz\cite{zhp} in the weak coupling approximation are: 
which initial states are least susceptible to loss of quantum 
coherence, and which are the likely end states.We are able to 
give precise answers to these questions without making the weak
coupling approximation due to a factorization property of the initial
state dependence of the exact solutions of the master equation.
We begin by proving the factorization theorem.

\vspace*{1.0cm}

\noindent {\bf II. Exact Solution of the Master Equation}  
\medskip

\noindent {\bf A. Oscillator case --} In the position representation we denote
\begin{equation}
 \langle x'|\rho(t)| x \rangle= \rho(R, r,t),
\end{equation}
where
\begin{equation}
R=\frac{x + x'}{2}, r= x-x'
\end{equation}
The master equation then becomes:
\begin{eqnarray}
\frac{\partial \rho (R,r,t)}{\partial t} &=& \Big [-\frac{i\hbar}{2m}
\bigl( \frac {\partial ^{2}}{\partial x^{2}} -
\frac{\partial ^{2}}{\partial x'^{2}} \bigr) - 
 \gamma (x-x ') \bigl( \frac {\partial}{\partial x} - 
\frac{\partial}
{\partial x'} \bigr)  
- \frac{D}{4 \hbar^{2}}(x-x')^{2}   \\ \nonumber
&& + \frac{m\omega^{2}}{2i\hbar}(x'^{2}-x^{2})\Big ]\rho (R,r,t) \\ \nonumber  
& = & -\Big [\frac{i\hbar}{m} \frac{\partial^{2}}{\partial r \partial
R} + 2 \gamma r 
\frac{ \partial}{\partial r} + \frac{Dr^{2}}{4 \hbar^{2}} + 
\frac{m\omega^{2}rR}{i\hbar} \Big ] \rho (R,r,t).
\end{eqnarray}
A Fourier transform with respect to $R$ reduces it to a first order partial
differential equation. Defining
\begin{eqnarray}
\rho (R,r,t)&=& \int \frac{dp dp'}{2\pi} \exp{(i(p'-p)R-i(p+p')r/2)}
\langle p'|\rho(t)|p \rangle \\ \nonumber 
&\equiv& \frac{1}{\sqrt{2\pi}} \int_{-\infty}^{\infty}dK e^{iKR}
 \tilde{\rho}(K,r,t), \hspace*{0.5 cm} K=p'-p,
\end{eqnarray}
we obtain
\begin{equation}
\frac{\partial \tilde{\rho}(K, r, t)}{\partial t} + \Big [ 
(2 \gamma r - \frac{\hbar K}{m})\frac{\partial}{\partial r} + 
\frac{m\omega^{2}r}{\hbar}\frac{\partial}{\partial K} + 
\frac{Dr^{2}}{4 \hbar^{2}} \Big] \tilde{\rho}(K, r, t)=0.
\end{equation}
To integrate this by Lagrange's method (of characteristics) note that on
a curve $K=K(s), t=t(s), r=r(s)$, the equation becomes
\begin{equation}
\frac{d\tilde{\rho}}{ds} + \frac{Dr^{2}}{4 \hbar^{2}}\tilde{\rho}=0,
\end{equation}
with
\begin{equation}
ds=\frac{dt}{1}=\frac{dr}{2 \gamma r - \frac{\hbar K}{m}}=\frac{dK}
{rm\omega^{2}/\hbar} =\frac{d\tilde{\rho}}{(-Dr^{2}/(4\hbar^{2}))\tilde{\rho}}
\end{equation} 
We readily obtain the three integrals
\begin{equation}
U_{\pm}=C_{\pm}, \hspace*{0.5cm} U_{3}=C_{3},
\end{equation}
where $C_{\pm}$ and $C_{3} $ are integration constants and 

\begin{eqnarray}
U_{\pm} &=& (K-\frac{r}{\lambda_{\pm}})\exp{(\frac{-\hbar t}
{m \lambda_{\pm}})}, \\  
U_{3}&=& \tilde {\rho}(K, r, t) \exp{ [ \frac{D} 
{32 m \hbar \left( \gamma^{2} - \omega^{2}\right)} \Big ( \lambda_{+}  
( K - \frac{r}{ \lambda_{+}} )^{2}} \\ \nonumber 
&& - \frac{ 2 \hbar }{ m \gamma } ( K - \frac{r}{ \lambda_{+}})
(K - \frac{r}{\lambda_{-}})  
+ \lambda_{-} ( K - \frac{r}{\lambda_{-}})^{2} \Big ) ],
\end{eqnarray}
and $\lambda_{\pm}$ are constants defined by
\begin{equation}
\lambda_{\pm}=\frac{\hbar}{m\omega^{2}}(\gamma \pm \sqrt{\gamma^{2}-
\omega^{2} } ). 
\end{equation}
If $(K, r, t)$ and $(K', r', 0)$ are points on the curve (15), Eqs. (16)
yield $K', r'$  in terms of $U_{\pm}$
\begin{equation}
K'=\frac{U_{+} \lambda_{+} - U_{-} \lambda_{-}}{(\lambda_{+} -
\lambda_{-})}, ~ r'= (U_{+} -
U_{-})\frac{\lambda_{+}\lambda_{-}}{\lambda_{+}-\lambda_{-}}. 
\end{equation}
Eq. (17) then yields the general solution,
\begin{equation}
\tilde {\rho}(K, r, t)=\tilde {\rho}(K', r', 0)\exp{(\alpha Z)},
\end{equation}
where
\begin{equation}
\alpha=\frac{D}{16 m^{2} (\gamma^{2}-\omega^{2})},
\end{equation}
and 
\begin{eqnarray}
Z&=&\frac{1}{\gamma}(K-\frac{r}{\lambda_{+}})
(K - \frac{r}{\lambda_{-}})(1-e^{-2 \gamma t}) \\ \nonumber
&& -\frac{m \lambda_{+}}{2 \hbar}(K-\frac{r}{\lambda_{+}})^{2}
(1-e^{-\frac{2 \hbar t}{m\lambda_{+}}})
\\ \nonumber
&& -\frac{m \lambda_{-}}{2 \hbar}(K-\frac{r}{\lambda_{-}})^{2}
(1-e^{-\frac{2 \hbar t}{m\lambda_{-}}}).
\end{eqnarray}
Eq. (20) is the {\underline {factorization theorem}}: the exact solution at an
arbitray time is equal to the initial reduced density operator with
shifted arguments $(K', r')$ times the function $\exp{(\alpha Z)}$ which
is independent of the initial conditions. It may be verified by direct 
substitution that the expression (20) solves Eq.(12). With
$\tilde{\rho}(K, r, t)$ known, a Fourier transform yields 
$\rho(R, r, t)$ explicitly.
\bigskip

\noindent {\bf B. Free Particle Case --} Starting from Eq. (10) with
$\omega=0$, or by taking the 
$\omega \rightarrow 0$ limit of Eqs(20)-(22), we obtain,
\begin{equation}
\tilde {\rho}(K, r, t)=\tilde {\rho}(K', r', 0)
\left(\exp{ -\frac{D}{16m^{2} \gamma^{2}} \Big [
K^{2} t + \frac{m (r-r')}{\hbar} 
\Big ( (r + r')\frac{m\gamma}{\hbar} +K \Big )
\Big ] }\right)
\end{equation}
where,
\begin{equation}
K'=K, r'=\frac{\hbar K}{2m\gamma} + (r-\frac{\hbar K}
{2 m \gamma})e^{-2 \gamma t}.
\end{equation}
The general solution (23) valid for arbitrary initial conditions 
agrees with the particular solution for an initial Gaussian wavepacket
obtained earlier in Ref [11] except that $\gamma$ should be replaced by
$4 \gamma$ in the solution quoted there.
\bigskip

\noindent {\bf III. Energy Basis Via Decoherence}
\medskip

\noindent {\bf A. Oscillator case --} The decoherence mechanism during the
system-bath interaction is known to suppress the off-diagonal elements
of the reduced density matrix of the system in an appropriate basis,
making all information on the system classically interpretable in that
basis. For simplified models where the self-Hamiltonian of the system
has either been ignored or considered co-diagonal with the interaction
Hamiltonian, the preferred basis in which the density matrix becomes
nearly diagonal has been believed to be the one that commutes with the
interaction Hamiltonian. In the Caldeira-Leggett Model studied here,
the coupling to the bath is a coordinate-coordinate coupling. Consider
the solutions, Eq(20) and Eq(23) for the oscillator and the free
particle, respectively, at long times ($\gamma t >>1$) [12]. For the
oscillator the solution at $t\rightarrow \infty$ for the overdamped
case ($\gamma >> \omega$) is:
\begin{equation}
\tilde {\rho}(K, r, \infty)= \tilde\rho (0,0,0)  
\exp{\Big [-\frac{D}{16m^{2}\omega^2
\gamma} \Big ( K^{2} + \frac{m^{2}
\omega^{2} r^{2}}{\hbar^{2}} \Big ) \Big]}.
\end{equation}
This limiting density matrix is actually independent of the initial
density matrix $\rho (0)$ provided that $\rho(0)$ is normalized,
because then 
\[
\tilde\rho(0,0,0) = {{\rm Tr} \rho(0) \over \sqrt{2\pi}} =
1/\sqrt{2\pi}.
\]
The final density matrix is not diagonal in the position basis or
momentum basis.  For example, in the position basis we find,
\begin{equation}
\langle x'|\rho(\infty)|x\rangle = A \exp(-\alpha_+(x^2 + x^{\prime
2}) - 2\alpha_- xx'),
\end{equation}
where
\begin{equation}
A = 4mw \sqrt{\gamma \over 4\pi D}, ~ \alpha_\pm = {m^2 w^2 \gamma
\over D} \pm {D \over 16\hbar^2 \gamma}.
\end{equation}
Since $\rho(\infty)$ is Hermitian, we seek an orthonormal complete set
of states $|\phi_n\rangle$ s.t.
\begin{equation}
\rho(\infty) |\phi_n\rangle = \lambda_n |\phi_n\rangle,
\end{equation}
and
\begin{equation}
\rho(\infty) = \sum_n \lambda_n |\phi_n\rangle \langle\phi_n|.
\end{equation}
Using the relation [13],
\begin{equation}
\int^\infty_{-\infty} e^{-(x-y)^2} H_n (\alpha x) dx = \sqrt{\pi} (1 -
\alpha^2)^{n/2} H_n \left({\alpha y \over \sqrt{1 - \alpha^2}}\right),
\end{equation}
we find the solutions $\phi_n$ of (28) to be,
\begin{equation}
\langle x|\phi_n\rangle = A_n e^{-x^2\left({mw \over 2\hbar}\right)}
H_n \left(x \sqrt{mw \over \hbar}\right),
\end{equation}
where $H_n(x)$ are the Hermite Polynomials, and $A_n$ are
normalization constants.  We also find the eigenvalues
\begin{equation}
\lambda_n = {8m \gamma \hbar w \over D + 4m \gamma \hbar w} \left({D -
4m \gamma \hbar w \over D + 4m \gamma \hbar w}\right)^n.
\end{equation}
Thus the final density matrix is diagonal in the basis of energy
eigenstates of the oscillator (Eq. (31)).  The eigenvalues $\lambda_n$
form a convergent geometric series due to the inequality (7) and yield
${\rm Tr} \rho (\infty) = 1$.

The Wigner function corresponding to the final density matrix is
easily seen to be of Gaussian form (in agreement with a theorem of
Tegmark and Shapiro [14]),
\begin{eqnarray}
W (x,p,t=\infty) &=& \int^\infty_{-\infty} {dy \over 2\pi \hbar}
\langle x - {y \over 2} |\rho(\infty)|x + {y \over 2}\rangle
e^{ipy/\hbar} \nonumber \\
&=& {4m\gamma w \over \pi D} \exp\left(-x^2 {4m^2 \gamma w^2 \over D}
- p^2 {4\gamma \over D}\right).
\end{eqnarray}
It yields the position and momentum uncertainties
\begin{equation}
\Delta x = \sqrt{D \over 8m^2 \gamma w^2}, ~ \Delta p = \sqrt{D \over
8\gamma}, 
\end{equation}
and
\begin{equation}
\Delta x \Delta p = {D \over 8\gamma m w} > {1\over 2} \hbar.
\end{equation}
The linear entropy corresponding to the final density matrix is,
\begin{equation}
S = {\rm Tr} (\rho(\infty) - \rho^2 (\infty)) = 1 - {4m \gamma \hbar w
\over D} > 0.
\end{equation}
Since the density matrix for $t \rightarrow \infty$ is independent of
the initial state, the final uncertainty product $\Delta x \Delta p$
and the entropy production (36) are also independent of the initial
state.  At intermediate times the exact density matrix (20) shows a
factorized dependence on the initial density matrix which we hope to
study in detail later.
\bigskip

\noindent {\bf B. Free Particle Case --} Eq. (23) shows that due to the
factor $\exp(-Dk^2 t/(16m^2 \gamma^2))$ on the right-hand side, the
density matrix is driven to a diagonal matrix in momentum space for $t
\rightarrow \infty$.  Further, the diagonal elements $(K = p' - p =
0)$ become 
\[
\tilde\rho (o,r,t) ~ \rarrow ~ {1 \over
\sqrt{2\pi}} \exp\left[-Dr^2/(16\gamma \hbar^2)\right].
\]
Since momentum being diagonal implies
energy being diagonal for a free particle, the energy basis emerges in
this example too as the preferred basis for $t \rightarrow \infty$.
\bigskip

\noindent {\bf Conclusion --} We have obtained exact solutions of the
Caldeira-Leggett Master  
equation for the reduced density matrix for an oscillator 
and for a free particle for arbitrary initial conditions in a compact
factorizable form. The solutions
for both cases studied show that the density matrix eventually 
diagonalizes in the energy basis at long times 
though the coupling to
the bath is via position. Our conclusion is in tune with the recent result
of Paz and Zurek \cite{paz} where they show that eigenstates of energy 
emerge as pointer states, but we do not use any 
weak coupling approximation.  The $t \rightarrow \infty$ form of our
Wigner function agrees with a  theorem of Tegmark and
Shapiro [14].  We hope to study later the intermediate time behavior of the
density matrix using the factorised exact solution obtained here.

\end{document}